\documentclass{aa}

\usepackage{amsmath,stmaryrd}
\usepackage[utf8x]{inputenc}
\usepackage[T1]{fontenc}
\usepackage{txfonts}
\usepackage{textcomp,wasysym}

\usepackage{natbib}
\bibpunct{(}{)}{;}{a}{}{,} 

\usepackage[french,english]{babel}

\usepackage[breaklinks]{hyperref}
\usepackage{graphicx}
\graphicspath{{fig/}}

\usepackage{xcolor}

\providecommand{\unit}[1]{\ensuremath{\:\mathrm{#1}}}

\newcommand{\ud}{\ensuremath\,\mathrm{d}}

\newcommand*\imax{\ensuremath{_{\textrm{max}}}}

\newcommand*{\lyeps}{Ly\,$\varepsilon$}
\newcommand*{\svi}{\ion{S}{vi}}
\newcommand*{\svii}{\ion{S}{vi}~93.3\,nm}
\newcommand*{\svij}{\ion{S}{vi}~94.4\,nm}
\newcommand*{\siviii}{\ion{Si}{viii}}

\newcommand*{\nE}{\ensuremath{n_{\textrm{e}}}}
\newcommand*{\nEav}{\ensuremath{\langle \nE \rangle}}
\newcommand*{\nEcav}{\ensuremath{\langle \nE^2 \rangle}}
\newcommand*{\nERMS}{\ensuremath{\nEav_{\text{RMS}}}}
\newcommand*{\nH}{\ensuremath{n_{\textrm{H}}}}
\newcommand*{\nS}{\ensuremath{n_{\textrm{S}}}}
\newcommand*{\nSvi}{\ensuremath{n_{\textrm{\svi}}}}
\newcommand*{\nSvii}{\ensuremath{n_{\textrm{\svi},i}}}

\newcommand*\moment[1]{(#1)}

\begin{document}
\selectlanguage{english}

\title{Electron density in the quiet solar coronal transition region
  from SoHO/SUMER measurements of \ion{S}{vi} line radiance and opacity}

\titlerunning{Density in the TR from \svi\ observations}

\author{E. Buchlin \and J.-C. Vial}

\offprints{E. Buchlin, \protect\url{eric.buchlin@ias.fr}}

\institute{Institut d'Astrophysique Spatiale, CNRS \& Université Paris
  Sud, Orsay, France.}

\date{Received\,:  / Revised date\,:
}

\abstract{
  The sharp temperature and density gradients in the coronal
  transition region are a challenge for models and observations.
}{
  We set out to get the average electron density $\nEav$
  in the region emitting the \svi\ lines.  We use two different
  techniques which allow to derive linearly-weighted (opacity method)
  and quadratically-weighted (Emission Measure method) electron
  density along the line-of-sight, in order to estimate a filling
  factor or to derive a thickness of the layer at the
    formation temperature of the lines.
}{
  We analyze SoHO/SUMER spectroscopic observations of the \svi\ lines,
  using the center-to-limb variations of radiance, the center-to-limb
  ratios of radiance and line width, and the radiance ratio of the
  $93.3$--$94.4 \unit{nm}$ doublet to derive the opacity. We also use
  the Emission Measure derived from radiance at disk center.
}{
  We get an opacity $\tau_0$ at \svii\ line center of the order of
  $0.05$.  The resulting average electron density $\langle \nE
  \rangle$, under simple assumptions concerning the emitting layer, is
  $2.4 \cdot 10^{16} \unit{m^{-3}}$ at $T = 2 \cdot 10^5 \unit{K}$.
  This value is higher than (and incompatible with) the values
  obtained from radiance measurements ($2 \cdot 10^{15}
  \unit{m^{-3}}$).  The last value leads to an electron pressure of
  $10^{-2} \unit{Pa}$.  Conversely, taking a classical value for the
  density leads to a too high value of the thickness of the emitting
  layer.
}{
  The pressure derived from the Emission Measure method compares well
  with previous determinations.  It implies a low opacity of
  $5\,10^{-3}$ to $10^{-2}$.  The fact that a direct derivation leads
  to a much higher opacity remains unexplained, despite tentative
  modeling of observational biases.  Further measurements, in \svi\
  and other lines emitted at a similar temperature, need to be
  done, and more realistic models of the transition region
    need to be used.
}



\maketitle

\section{Introduction}

In the simplest description of the solar atmosphere, where it is
considered as a series of concentric spherical layers of plasma at
different densities and temperatures, the transition region (hereafter
TR) between the chromosphere and the corona is the thin interface
between the high-density and low-temperature chromosphere (a few
$10^{16} \unit{m^{-3}}$ hydrogen density at about $10^4 \unit{K}$) and
the low-density and high-temperature corona (about $10^{14}
\unit{m^{-3}}$ at $10^6 \unit{K}$).  The variation of temperature $T$
and electron number density \nE\ has been mostly derived from the
modelling of this transition region, where radiative losses are
balanced by thermal conduction \citep[e.g.][]{mariska93book,avrett08}.

Measurements of the electron density usually rely either on estimation
of the Emission Measure or on line ratios.  On one hand, using
absolute line radiances, the Emission Measure (EM) and Differential
Emission Measure (DEM) techniques provide $\nEcav$ at the formation
temperature of a line (or as a function of temperature if several
lines covering some range of temperatures are measured).  On the other
hand, the technique of line radiance ratios provides a wealth of
values of $\nE$ \citep{mason98lnp} with the assumption of uniform
density along the line-of-sight, and with an accuracy limited by the
accuracy of the two respective radiance measurements: typically, a
15\,\% uncertainty on line radiance measurement leads to 30\,\%
uncertainty on the line ratio and then to about a factor 3 uncertainty
on the density.  However, for a given pair of lines, this technique
only works in a limited range of densities.  Let us add that the
accuracy is also limited by the precision of atomic physics data.

Here we propose to use also the concept of opacity (or optical
thickness) in order to derive the population of the low (actually the
ground) level $i$ of a given transition $i \rightarrow j$, and then
the electron density.  At a given wavelength, the opacity of a column
of plasma corresponds indeed to the sum of the absorption coefficients
of photons by the individual ions in the column.  The opacity can be
derived by different complementary techniques \citep{dumont83} if many
measurements are available with spatial (preferably center-to-limb)
information.  This is the case in a full-Sun observations program by
the SoHO/SUMER UV spectro-imager \citep{sumer, peter99, peter99b} run
in 1996.  In particular, thanks to a specific ``compressed'' mode, a
unique dataset of 36 full-Sun observations in \svi\ lines has been
obtained; this makes possible to derive at the same time $\nEav$ from
opacity measurements and $\nEcav$ from line radiance measurements (via
the EM).

We have already used this data set in order to get properties of
turbulence in the TR \citep{buc06}.  Note that here, contrary
to \citet{peter99,peter99b,buc06}, we are not interested in the
resolved directed velocities or in the non-thermal velocities but in
the line radiances, peak spectral radiances and widths.  Also note
that, along with the modelling work of \citet{avrett08}, we do
not distinguish network and internetwork (anyway a difficult task at
the limb) and aim at a precise determination of the properties of an
average TR.

This paper is organized as follows: we first present the data set we
use, then we determine opacities and radiances of \svii, we get two
determinations of density in the region emitting the \svii\ line, we
discuss the disagreement between the two determinations (especially
possible biases), and we conclude.

\section{Data}
\label{sec:data}

\subsection{Data sets}

We use the data from a SoHO/SUMER full-Sun observation program in
\svii, \svij\ and \lyeps\ designed by Philippe Lemaire.  The
spectra, obtained with detector A of SUMER and an exposure time of
$3\unit{s}$, were not sent to the ground (except for context spectra)
but 5 parameters (``moments'') of 3 lines were computed on-board for
each position on the Sun:
\begin{itemize}
\item \moment{1} peak spectral radiance, \moment{2} Doppler shift, and
  \moment {3} width of the line \svii,
\item \moment{4} line radiance (integrated spectral radiance) of the line
  \lyeps 93.8\,nm,
\item \moment{5} line radiance of the line \svij.  It must be noted
  that this line is likely to be blended with \siviii.
\end{itemize}
The detailed characteristics of these lines can be found in
Table \ref{tab:lines}.  A list of the 36 observations of this program
run throughout year 1996, close to solar minimum, can be found in
Table 1 of \cite{buc06}.
These original data constitute the main data set we use in this
paper, hereafter DS1.  They are complemented by a set of 22 context
observations from the same observation program, that we use when we
need the full profiles of the spectral lines close to disk center: the
full SUMER detector ($1024\times360$ pixels) has been recorded at a
given position on the Sun at less than $40\unit{arcsec}$ from disk
center and with an exposure time of $300\unit{s}$.  This data is
calibrated using the Solar Software procedure
\verb'sum_read_corr_fits' (including correction of the flat field, as
measured on 23 September 1996, and of distortion), and it will hereafter
be referred to as DS2.

\begin{table*}[tp]
  \begin{minipage}[t]{\linewidth}
    \caption{Spectral lines present in the data sets, with parameters
      computed by CHIANTI and given by previous observations.}
    \centering
    \renewcommand{\footnoterule}{} 
    \begin{tabular}{ll|lll|ll}
      \hline \hline
      &            & \multicolumn{3}{c|}{CHIANTI\footnote{Using the
          ``Arnaud \& Raymond'' ionization fractions file, the ``Sun
          coronal'' abundance file and the ``Quiet Sun'' DEM file. 
          CHIANTI does not include data for the Hydrogen lines
          (\lyeps\ in particular).}} &
      \multicolumn{2}{c}{\cite{cur01}} \\
      Ion & Transition $j \rightarrow i$ & $\log T_{\rm max}$ (K) &
      Wavelength (\AA) & Radiance\footnote{Radiances 
        are given in $\unit{W\,m^{-2} sr^{-1}}$, and peak spectral
        radiances are given in $\unit{W\,m^{-2}sr^{-1}nm^{-1}}$.} &
      Wavelength (\AA) &
      Peak radiance$^b$ \\ \hline
      \svi & $ 2\mathrm{p}^6\;3\mathrm{p}\;{}^2\mathrm{P}_{3/2}
      \rightarrow 2\mathrm{p}^6\;3\mathrm{s}\;{}^2\mathrm{S}_{1/2}$ &
      $5.3$ & $933.3800$ & $3.81\cdot10^{-3}$ & $933.40$ & $0.57$ \\
      \lyeps & $ 6\mathrm{p}\;{}^2\mathrm{P}_{3/2} \rightarrow
      1\mathrm{s}\;{}^2\mathrm{S}_{1/2}$ & --- & --- & --- & $937.80$ &
      $1.07$ \\ 
      \siviii & $ 2\mathrm{s}^2\;2\mathrm{p}^3\;{}^2\mathrm{P}_{3/2}
      \rightarrow 2\mathrm{s}^2\;2\mathrm{p}^3\;{}^4\mathrm{S}_{3/2}$ &
      $5.9$ &
      $944.4670$ & $5.24\cdot10^{-3}$ & $944.34$ & $0.14$ \\ 
      \svi & $ 2\mathrm{p}^6\;3\mathrm{p}\;{}^2\mathrm{P}_{1/2}
      \rightarrow 2\mathrm{p}^6\;3\mathrm{s}\;{}^2\mathrm{S}_{1/2}$ &
      $5.3$ & $944.5240$ & 
      $1.91\cdot10^{-3}$ & $944.55$ & $0.29$ \\
      \hline
    \end{tabular}
    \label{tab:lines}
  \end{minipage}
\end{table*}

\begin{figure}[tp]
  \centering
  \includegraphics[width=\linewidth]{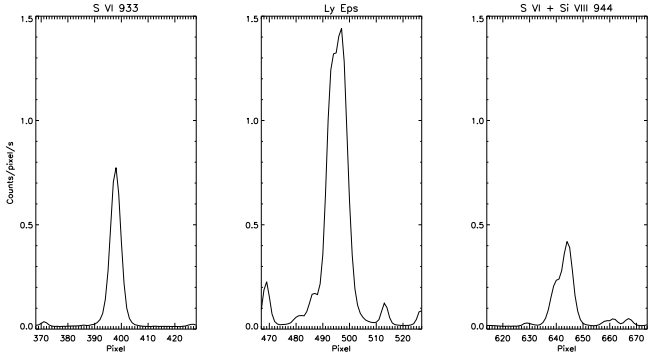}
  \caption{Raw line profiles from the context spectrum taken on 4 May
    1996 at 07:32 UT at disk center with an exposure time of
    $300\unit{s}$. The profiles are averaged over pixels 50 to 299
    along the slit ($1\times 300\unit{arcsec}$, detector A), with no
    prior destretching of the data.}
  \label{fig:profile}
\end{figure}

\subsection{Averages of the data as a function of distance to disk
  center}

\begin{figure}[tp]
  \centering
  \includegraphics[width=\linewidth]{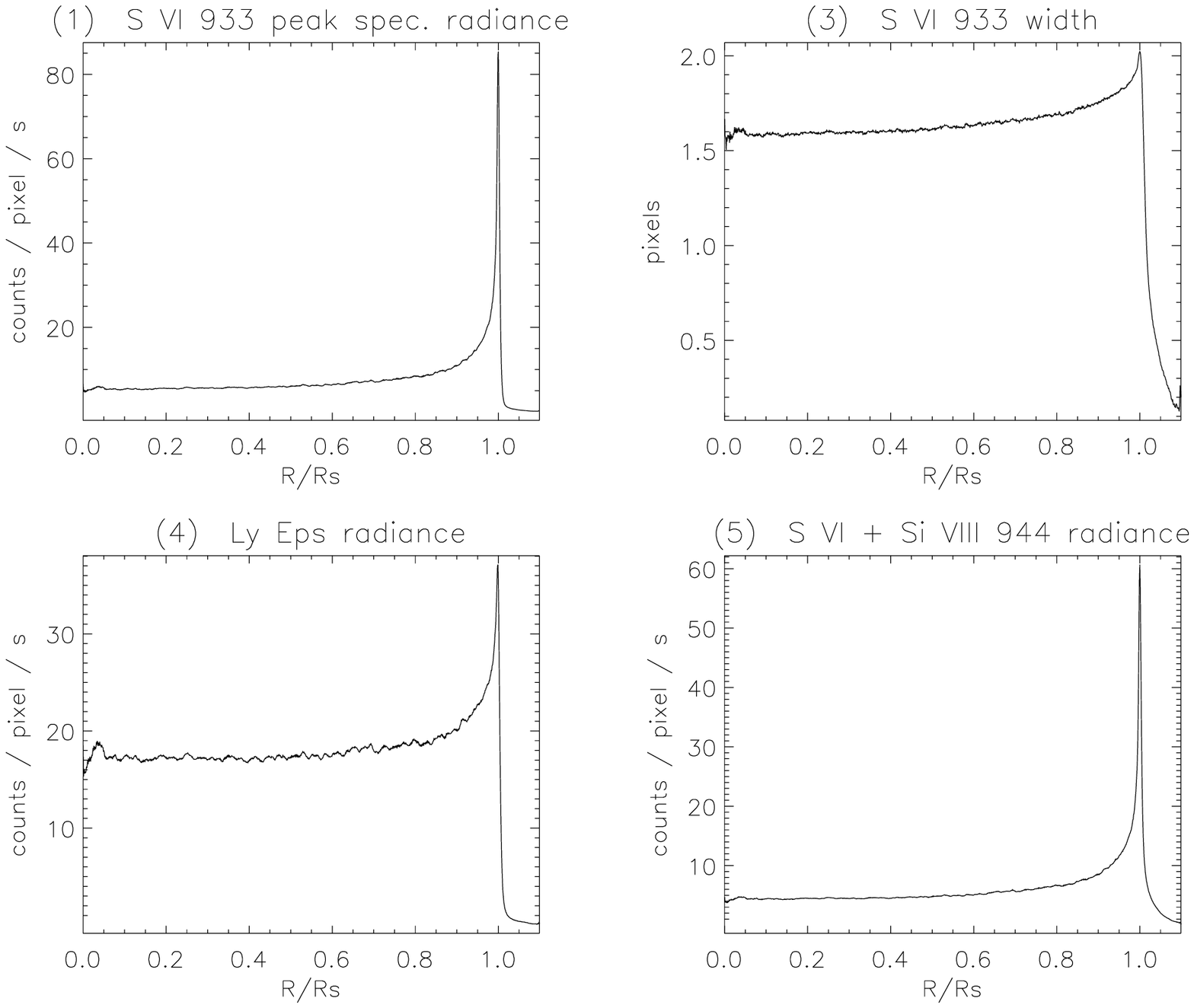}\\
  \includegraphics[width=\linewidth]{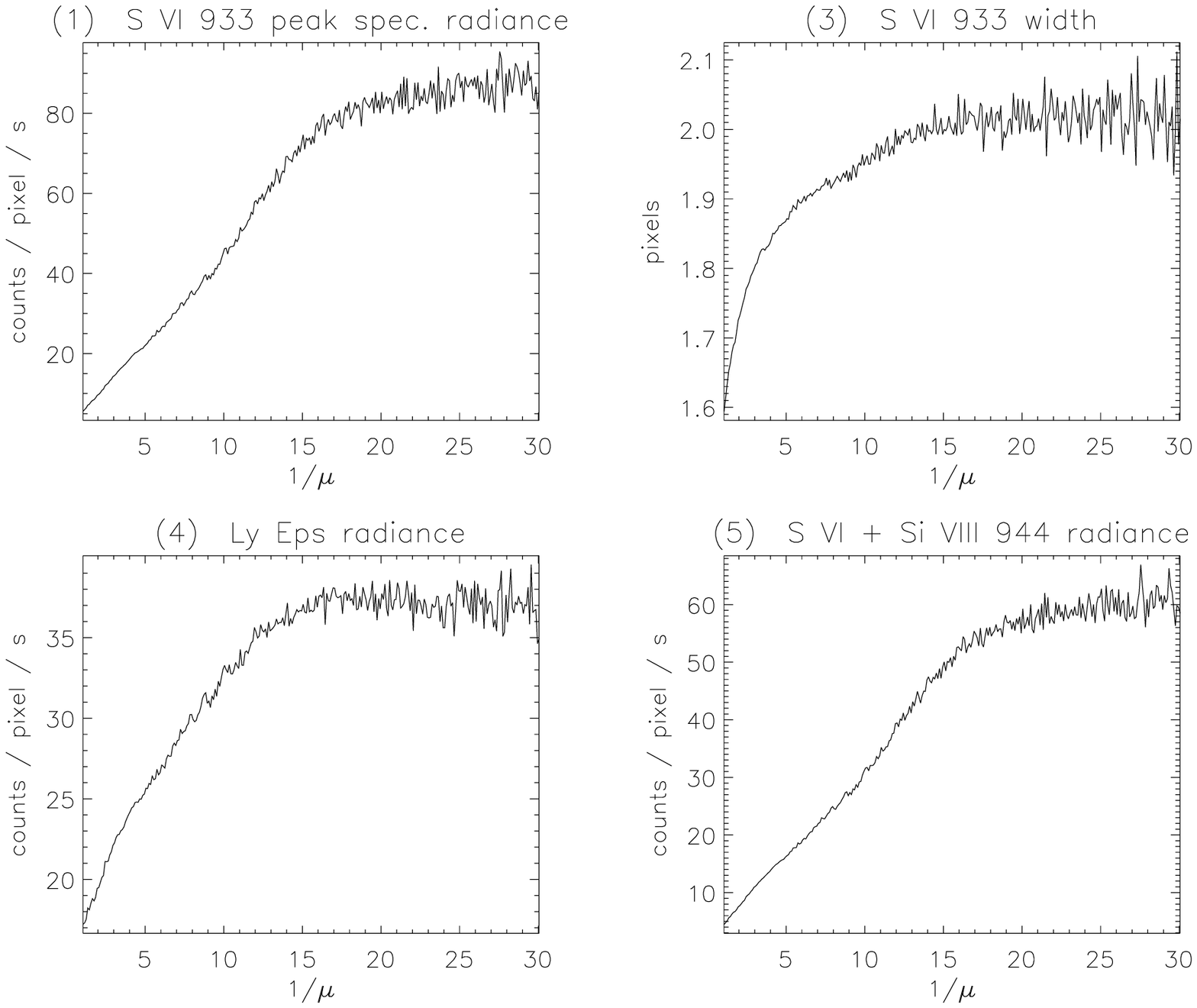}
  \caption{Average of the data as a function of $r /
    R_{\astrosun}$ (top panels) and as a function of $1 / \mu$ (bottom
    panels).}
  \label{fig:muprof}
\end{figure}

In order to obtain averages of the radiances in data set DS1 as a
function of the radial distance $r$ to the disk center, and as a
function of $\mu$, the cosine of the angle between the normal to the
solar ``surface'' and the line-of-sight, we apply the following method,
assuming that the Sun is spherical:
\begin{itemize}
\item We detect the limb automatically by finding the maximum of the
  \svii\ radiance at each solar-$y$ position in two detection windows
  in the solar-$x$ direction, corresponding to the approximate
  expected position of the limb.  This means that the limb is
  found in a TR line and is actually approximately $3\unit{arcsec}$
  above the photosphere.  However, this is the relevant limb position
  for the geometry of the \svii\ emission region.
\item We fit these limb positions to arcs of a circle described by
  $x(y)$ functions, and we get the real position $(a,b)$ of the solar
  disk center in solar coordinates $(x,y)$ given by SUMER, and the
  solar radius $R_{\astrosun}$ (this changes as a function of the time
  of year due to the eccentricity of SoHO's orbit around the Sun).
  The solar radius is evaluated for the observed wavelength of
  $93.3\unit{nm}$.
\item We choose to exclude zones corresponding to active regions, as
  the aim of this paper is to obtain properties of the TR in the Quiet
  Sun.
\item For each of the remaining pixels, we get values of the radial
  distance $r = \sqrt{(x-a)^2 + (y-b)^2}$ to disk center and of $\mu =
  \sqrt{1-(r / R_{\astrosun})^2}$.
\item We compute the averages of each moment (radiances and widths) in
  bins of $r/R_{\astrosun}$ and in bins of $1/\mu$.
\end{itemize}
The resulting averages as a function of $r/R_{\astrosun}$ and of
$1/\mu$ are plotted in Fig. \ref{fig:muprof} (except for the \svii\
Doppler shift, which will not be used in this paper).  The radiances
are approximately linear functions of $1/\mu$ for small $1/\mu$, as
expected from optically thin lines in a plane-parallel geometry. Such
a behavior actually validates the consideration of a ``mean''
plane-parallel transition region, at least for $1/\mu < 10$ or $\theta
< 84$\textdegree.

\section{Determination of opacities}
\label{sec:opa}

\subsection{Using center-to-limb variations}
\label{sec:methoda}

We follow here the method A proposed by \citet{dumont83}. Assuming
that the TR is spherically symmetric and that it can be considered as
plane-parallel when not seen too close to the limb, that the lines are
optically thin, and that the source function $S$ is constant in the
region where the line is formed\footnote{We release this assumption in
  Sec.~\ref{sec:disc}.}, the spectral radiance is:
\begin{equation}
  \label{eq:ivar}
  I_0(\mu) = S (1 - \exp (-\tau_0 / \mu))
\end{equation}
where the subscript $0$ is for the line center and $\tau$ is the
opacity of the emitting layer at disk center.  Then:
\begin{equation}
 \label{eq:iprof}
  I_0(\mu) = I_0(1) \frac{1 - \exp (-\tau_0 / \mu)}{1 - \exp
    (-\tau_0)}
\end{equation}
and a fit of the observed $I_0(\mu)$ by this function, with $I_0(1)$
and $\tau_0$ as parameters\footnote{Note that, contrary to
  \citet{dumont83}, we take $I_0(1)$ as an additional parameter.  This
  is because by doing so, we avoid the sensitivity of $I_0(1)$ to
  structures close to disk center, and because the first data bin
  \emph{starts} at $1/\mu=1$ instead of being centered on $1/\mu=1$},
gives an estimate of $\tau_0$.

For the lines for which only the line radiance $E$ is known (\svij\
and \lyeps), we need to fit this function, with $\tau_0$ and $E(1)$ as
parameters\footnote{We take here $E(1)$ as a parameter for the same
  reason as we did before for $I_0(1)$.}:
\begin{equation}
  \label{eq:eprof}
  E(\mu) = E(1) \frac{\int_{\mathbb{R}} \left( 1 - \exp
      \left(-\frac{\tau_0}\mu \, e^{-u^2}\right) \right) \,
    \text{d}u}{\int_{\mathbb{R}}
    \left( 1 - \exp
      \left(-\tau_0 \, e^{-u^2}\right) \right) \, \text{d}u}
\end{equation}
This expression comes from \citet{dumont83} and assumes a Doppler
absorption profile $\exp (-u^2)$ with $u = \Delta \lambda / \Delta
\lambda_D$.  Here, contrary to the case of the peak spectral radiance
ratio, the function and its derivative with respect to $\tau_0$ and
$E(1)$ cannot be computed analytically anymore, and we need to
estimate them numerically; this is done by a fast method, using a
Taylor expansion of the outermost exponential of both the numerator
and denominator of Eq. (\ref{eq:eprof}).

\begin{figure}[tp]
  \centering
  \includegraphics[width=\linewidth]{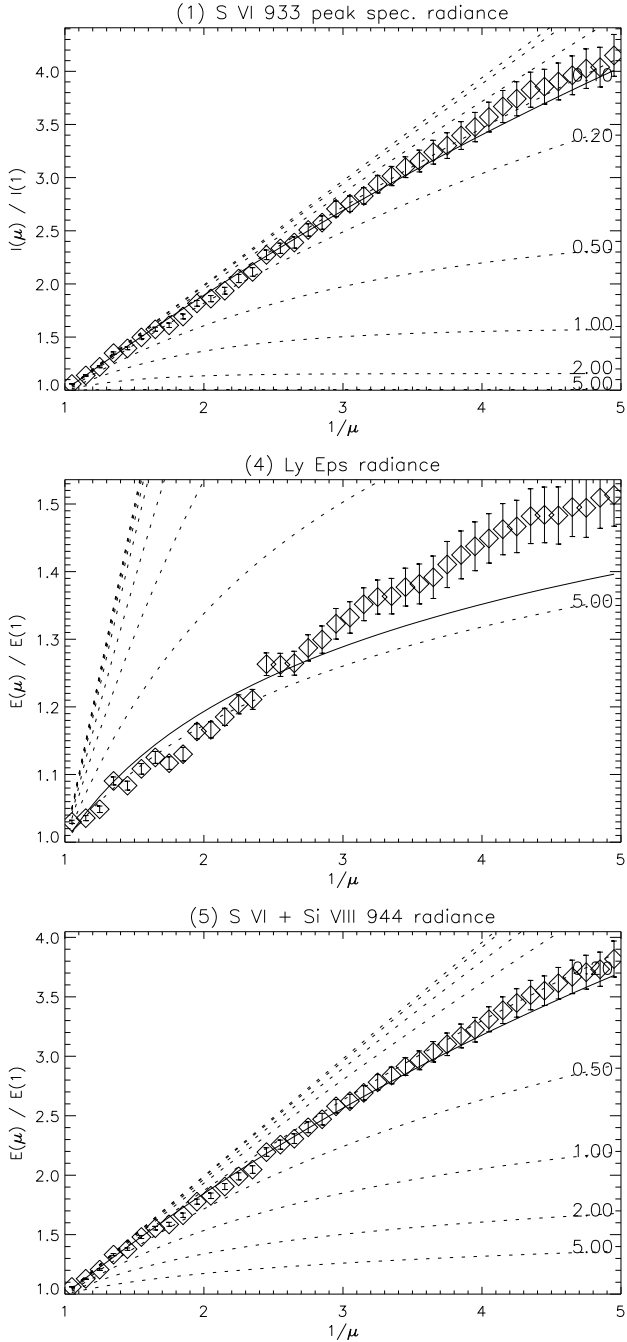}
  \caption{Diamonds: average profiles of the radiance data (moments 1,
    4 and 5) as a function of $1/\mu$, normalized to their values at
    disk center. Dotted lines: theoretical profiles for different
    values of $\tau_0$.  Plain lines: fits of the theoretical profiles
    to the data, giving the values for $\tau_0$: $0.113$ for
    \moment{1} and $0.244$ for \moment{5}.  The fit for \lyeps\ is bad
    because this line is optically thick.}
  \label{fig:muproffit}
\end{figure}

These theoretical functions of $\mu$ are then plotted for different
values of the parameter $\tau_0$ over the observations in
Fig. \ref{fig:muproffit}, for all three lines (either for the peak
spectral radiance or the line radiance, depending on the data).  We
have performed a non-linear least-squares fit using the
Levenberg-Marquardt algorithm as implemented in the Interactive Data
Language (IDL); it gives the parameter $\tau_0$.  The uncertainties on
each point of the $E(\mu)$ or $I(\mu)$ functions (an average on $N_d$
pixels) that we take as input to the fitting procedure come mainly
from the possible presence of coherent structures such as bright
points: the number of such possible structures is of order $N_d /
N_s$, where $N_s$ is the size of such a structures (we take $N_s=100$
pixels), and then the uncertainty on $I$ or $E$ is $\sigma / \sqrt{N_d
  / N_s}$ where $\sigma$ is the standard deviation of the data points
(in each pixel of a $1 / \mu$ bin).  Compared to this uncertainty, the
photon noise is
negligible.

The results of the fits on the interval $1/\mu \in [1,5]$ are shown in
Fig. \ref{fig:muproffit}: as far as $\tau_0$ is concerned, they are
$0.113$ for moment \moment{1} (\svii\ peak spectral radiance) and
$0.244$ for moment \moment{5} (\svij\ radiance, blended with \siviii).
The approximations we used in writing Eq. (\ref{eq:ivar}) are not
valid for the optically thick \lyeps\ line, hence the bad fit.  On the
other hand, these approximations are valid for both the \svi\ lines,
as long as $1/\mu$ is small enough.  For large $1 / \mu$ there is an
additional uncertainty resulting from the determination of the limb.

These results are somewhat sensitive to the limb fitting: a $10^{-3}$
relative error in the determination of the solar radius leads to a $7
\; 10^{-2}$ relative error on $\tau_0$.  As $10^{-3}$ is a conservative
upper limit of the error on the radius from the limb fitting, we can
consider that $7 \; 10^{-2}$ is a conservative estimate of the
relative error on $\tau_0$ resulting from the limb fitting.

\subsection{Using center-to-limb ratios of \svii\ width and radiance}
\label{sec:methodb}

The variation with position of the \svii\ line width (see
Fig. \ref{fig:muprof}) can be interpreted as an opacity saturation of
the \svii\ line at the limb, and then method B of \cite{dumont83} can
be applied.  This method relies on the measurement of the ratio $d =
\Delta \lambda^*_l / \Delta \lambda^*_c$ of the FWHM at the limb and
at the disk center: the optical thickness at line center $t_0$ at the
limb is given by solving
\begin{equation}
2 \left(1 - \exp \left(-t_0 \, e^{-d^2 \ln 2}\right)\right) = 1 -
\exp (-t_0)
\end{equation}
(this is Eq. 4 of \citealt{dumont83} where a sign error has been
corrected) and then the opacity at line center $\tau_0$ is
given by solving
\begin{equation}
  \frac{I_0 (\mu=1)}{I_0 (\mu = 0)} = \frac{1 - \exp(-\tau_0)}{1 -
    \exp(-t_0)}
\end{equation}

Using the full-Sun \svii\ compressed data set DS1\footnote{Although
  not obvious from the data headers, moment \moment{3} corresponds to
  the deconvoluted FWHM of \svii, as is confirmed by a comparison with
  the width obtained from the full profiles in data set DS2 and
  deconvoluted using the Solar Software procedure
  \texttt{con\string_width\string_4}.}, we find that the ratio $d$ is
$1.274$ and then $t_0$ is $1.53$.  Finally, we use the \svii\ peak
spectral radiance ratio $I_0(\mu=1)/I_0(\mu=0) = 0.062$ to
get $\tau_0 = 0.05$.

\subsection{Using the \svi\ 94.4 -- 93.3 line ratio }
\label{sec:methodc}

The theoretical dependence of the \svi\ 94.4 -- 93.3 peak radiance
line ratio as a function of the line opacities and source functions is: 
\begin{equation}
  K = \frac{I_{0,933}}{2 \, I_{0,944}} = \frac{S_{933}}{2 \, S_{944}} \frac{1 -
  \exp (-\tau_{0,933})}{1 - \exp (-\tau_{0,944})} 
\end{equation}
For this doublet, we assume $S_{933} = S_{944}$ and $\tau_{0,933} = 2
\tau_{0,944}$ (because the oscillator strengths are in the proportion
$f_{933} = 2 f_{944}$). Then $K$ reduces to
\begin{equation}
  K = \frac12 \frac{1 -
  \exp (-\tau_{0,933})}{1 - \exp (-\tau_{0,933}/2)} = \frac{1 +
\exp (-\tau_{0,933} / 2)}{2}
\end{equation}
and we get $\tau_{0,933}$ from the observed value of $K$:
\begin{equation}
  \tau_{0,933} = -2 \ln (2K - 1) 
\end{equation}

The difficulty comes from the \svij\ blend with the \siviii\ line. In
order to remove this blend, we have analyzed the line profiles
available in data set DS2.  After averaging the line profiles over the
60 central pixels along the slit, we have fitted the \svii\ line by a
Gaussian with uniform background and the \svij\ line blend by two
Gaussians with uniform background.  We have then computed the Gaussian
amplitude from these fits for both \svi\ lines, and this gives
$I_{0,933}$ and $I_{0,944}$, and then $K$, that we average over all
observations.  From this method we get $\tau_{0,933} = 0.089$.

The same kind of method could in theory be used for the \svi\ 94.4 --
93.3 line radiance ratio
\begin{equation}
  \label{eq:ke}
  K = \frac{E_{933}}{2 \, E_{944}} = \frac{S_{933}}{2 \, S_{944}}
  \frac{\int_{\mathbb{R}} \left(1 -
      \exp \left(-\tau_{0,933} \, e^{-u^2}\right)\right) \ud
    u}{\int_{\mathbb{R}} \left(1 - \exp \left(-\tau_{0,944}
        \, e^{-u^2}\right) \right) \ud u}
\end{equation}
with, again, $S_{933} = S_{944}$ and $\tau_{0,933} = 2 \tau_{0,944}$.
As for method A, the integral makes it necessary to invert this
function of $\tau_{0,933}$ numerically, in order to recover
$\tau_{0,933}$ for a given observed value of $K$.  As $K$ is
decreasing as a function of $\tau_{0,933}$, this is possible by a
simple dichotomy.  However, the average $K$ from the observations is
greater than $1$, which makes it impossible to invert the function and
get a value for $\tau_0$.

\subsection{Discussion on opacity determination}

It is clear that the three methods provide different values of the
opacity at disk center. We confirm the result of \citet{dumont83},
obtained in different lines, by which the method of center-to-limb
ratios of width and radiance (Sec. \ref{sec:methodb}, or method B in
\citealt{dumont83}) provides the smallest value of the opacity.  As
mentioned by these authors, the center-to-limb variations method
(Sec. \ref{sec:methoda}, or method A) overestimates the opacity for
different reasons described in \citet{dumont83}, among which the
curvature of the layers close to the limb and their roughness.  The
method of line ratios (Sec.\ref{sec:methodc}, or method C) also
provides larger values of the opacity, although free from geometrical
assumptions; \citet{dumont83} interpret them as resulting from a
difference between the source functions of the lines of the doublet.

This does not mean that there are no additional biases.  For instance,
we have adopted a constant Doppler width from center to limb; actually
this is not correct since at the limb the observed layer is at higher
altitude, where the temperature and turbulence are higher than in the
emitting layer as viewed at disk center.  Consequently, the excessive
line width is wrongly interpreted as only an opacity effect.  However,
it seems improbable that a $27.4\%$ increase of Doppler width from
center to limb can be entirely interpreted in terms of temperature
(because of the square-root temperature variation of Doppler width)
and turbulence (as the emitting layer is --- a posteriori ---
optically not very thick).

\section{First estimates of densities}
\label{sec:den}

\newcommand*\Abund{\mathop{\mathrm{Abund}}}

\subsection{Densities using the opacities}
\label{sec:ne}

The line-of-sight opacity at line center of the \svii\ line is given by
\begin{equation}
  \label{eq:tau0}
  \tau_0 = \int k_{\nu_0} \, \nSvii(s) \, \textrm{d}s
\end{equation}
where the integration is along the line-of-sight.  The variable
\nSvii\ is the numerical density of \svi\ in its level $i$, which can be
written as
\begin{equation}
  \label{eq:ns}
  \nSvii = \frac\nSvii\nSvi \frac{\nSvi}{\nS} \Abund(\textrm{S})
  \frac{\nH}{\nE} \nE
\end{equation}
where $\Abund(\textrm{S}) = \nS / \nH$ is the Sulfur abundance in the
corona ($10^{-4.73}$ according to the CHIANTI database,
\citealt{chianti,landie06}), $\nSvii/\nSvi$ is the proportion of \svi\
at level $i$, $\nSvi/\nS$ is the ionization fraction (known as a
function of temperature) and $\nH / \nE = 0.83$ is constant in a fully
ionized medium as the upper transition region.  In this work $i$ is
the ground state $i=1$, and as in this region $n_{\textrm{\svi},1} /
\nSvi$ is very close to $1$, we will drop this term from now.  The
variable $k_{\nu_0}$ is the absorption coefficient at line center
frequency $\nu_0$ for each \svi\ ion, given by:
\begin{equation}
  \label{eq:knu0}
  k_{\nu_0} = \frac{h\nu_0}{4\pi} B_{ij} \frac1{\sqrt{\pi} \, \Delta\nu_D}
\end{equation}
where $B_{ij}$ is the Einstein absorption coefficient for the
transition $i \rightarrow j$ (i.e.,
$2\mathrm{p}^6\;3\mathrm{s}\;{}^2\mathrm{S}_{1/2} \rightarrow
2\mathrm{p}^6\;3\mathrm{p}\;{}^2\mathrm{P}_{3/2}$) at $\lambda_0 =
93.3\unit{nm}$ and integration over a Gaussian Doppler shift
distribution has been done ($\Delta\nu_D$ is the Doppler width in
frequency).  Using:
\begin{equation}
  B_{ij} = \frac{g_j}{g_i} B_{ji} = \frac{g_j}{g_i} \frac{A_{ji}} {2 h
  \nu_0^3 / c^2}
\end{equation}
with $g_j / g_i = 2$ and $\lambda_0 = c / \nu_0$, this gives:
\begin{equation}
  k_{\nu_0} = \frac{\lambda_0^4 A_{ji}}{4\pi^{3/2} c \, \Delta\lambda_D}
\end{equation}

Finally, for an emitting layer of thickness $\Delta s$ and average electron
density $\nEav$, we have:
\begin{equation}
  \label{eq:tau0ne}
  \tau_0 = \frac{\lambda_0^4 A_{ji}}{4\pi^{3/2} c \, \Delta\lambda_D}
  \frac{\nSvi}{\nS} \Abund(\textrm{S})
  \frac{\nH}{\nE} \nEav \, \Delta s
\end{equation}
Taking $\tau_0 = 0.05$, we get $\nEav \, \Delta s = 4.9 \cdot 10^{21}
\unit{m^{-2}}$. Then, with $\Delta s = 206\unit{km}$ (the altitude
interval corresponding to the FWHM of the \svii\ contribution function
$G(T)$ as computed by CHIANTI), this gives $\nEav = 2.4 \cdot 10^{16}
\unit{m^{-3}}$.

\subsection{Squared densities using the contribution function}
\label{sec:ne2}

The average \svii\ line radiance at disk center obtained from data set
DS2 (excluding the 5\% higher values which are considered not to be
part of the quiet Sun) is $E = 1.4\cdot10^{-2}\unit{W\,m^{-2}
  sr^{-1}}$ (to be compared to the value $3.81\cdot10^{-3}$ given by
CHIANTI with a Quiet Sun DEM --- see Table~\ref{tab:lines}).  This can
be used to estimate $\nEcav \, \Delta s$ in the emitting region of
thickness $\Delta s$, as
\begin{equation}
  \label{eq:e}
  E = \int G(T(s)) \, \nE^2(s) \; \ud s \approx G(\langle T\rangle) \,
  \nEcav \, \Delta s 
\end{equation}
where $G(T)$ is the contribution function and the integral is on the
line-of-sight and where we have made the assumption that $\tau_0 \ll
1$.  We take the average temperature in the emitting region to be
$\langle T \rangle = T_{\rm max} = 10^{5.3} \unit{K}$, and, for
densities of the order of $10^{16} \unit{m^{-3}}$, the \verb'gofnt'
function of CHIANTI gives $G(\langle T\rangle) = 1.8 \cdot 10^{-37}
\unit{W \, m^3 sr^{-1}}$.  We finally get
\begin{equation}
  \nEcav \, \Delta s = 8.4 \cdot 10^{35} \unit{m^{-5}}
\end{equation}

With again $\Delta s = 206 \unit{km}$, we get $\nERMS = 2.0 \cdot
10^{15} \unit{m^{-3}}$.  Assuming an uncertainty of $20\%$ on $E$, the
uncertainty on $\nERMS$ would be $10\%$ for a given $\Delta s$.

\section{Discussion of biases in the method}
\label{sec:disc}

One of our aims when starting this work was to determine a filling
factor\footnote{We explain this definition of the filling factor in
  Appendix~\ref{sec:ff}.}
\begin{equation}
  \label{eq:ff}
  f = \frac{\nEav^2}{\nEcav}
\end{equation}
in the \svi-emitting region.  This initial objective needs to be
revised, since we get $f=144$, an impossible value as it is more than
$1$.  Our values of densities can be compared to the density at $\log
T = 5.3$ in the \citet{avrett08} model ($1.7 \cdot 10^{15}
\unit{m^{-3}}$): our value of $\nEav$ is an order of magnitude higher,
while $\nERMS =\sqrt{\nEcav}$ is about the same (while it should be
higher than $\nEav$).  Our value of intensity is compatible with
average values from other sources, such as \citet{delzannag01b} (see
their Fig. 1).

Given the same measurements of $\tau_0$ and $E$, one can
  instead start from the assumption of a filling factor $f\in[0,1]$
  and deduce $\Delta s$:
  \begin{equation}
    \label{eq:dsff}
    \Delta s = \frac1f \frac{(\nEav \, \Delta s)^2}{\nEcav \, \Delta s}
  \end{equation}
  where the numerator and denominator of the second fraction are
  deduced from Eq.~(\ref{eq:tau0ne}) and~(\ref{eq:e}) respectively.
  With the values from Sec.~\ref{sec:den}, this gives $\Delta s >
  29\unit{Mm}=0.04 R_{\astrosun}$, a value much larger than
  expected.

In any case, there seems to be some inconsistencies
around $\log T = 5.3$ between our new observations of opacities on one
hand, and transition region models and observations of intensities on
the other hand.  We propose now to discuss the possible sources of
these discrepancies, while releasing, when needed, some of the
simplistic assumptions we have made until now.

\subsection{Assumption of a uniform emitting layer}
\label{sec:biasuniform}

\subsubsection{Bias due to this assumption}

When computing the average densities from the \svii\ opacity and
radiance, we have assumed a uniform emitting layer at the temperature
of maximum emission and of thickness $\Delta s$ given by the width of
contribution function $G(T)$.  However, the different dependences in
the electron density of Eqs. (\ref{eq:tau0}) and (\ref{eq:e}) --- the
first is linear while the second is quadratic --- means that the slope
of the $\nE(s)$ function affects differently the weights on the
integrals of Eqs. (\ref{eq:tau0}) and (\ref{eq:e}): a bias, different
for $\tau_0$ and $E$, can be expected, and here we explore this effect
starting from the \citet{avrett08} model, which has the merit of
giving average profiles of temperature and density (among other
variables) as a function of altitude $s$.

\paragraph{Opacity.} Using the \citet{avrett08} profiles and atomic
physics data, we get $\tau_0 = 0.008$.  Then, using the same
simplistic method as for observations (still with a uniform layer of
thickness $\Delta s = 206 \unit{km}$), we obtain $\nEav
= 2.4 \cdot 10^{15} \unit{m^{-3}}$, a value only 40\% higher
than the density at $\log T = 5.3$ in this model ($1.7 \cdot 10^{15}
\unit{m^{-3}}$).

\paragraph{Radiance.} Using the same \citet{avrett08} profiles and the
CHIANTI contribution function $G(T)$, we get $E = 1.3 \cdot
10^{-2}\unit{W\,m^{-2} sr^{-1}}$.

Then, using the same simplistic method as for observations, we obtain
$\nERMS = 1.9 \cdot 10^{15}\unit{m^{-3}}$, a
value 12\% higher than the density at $\log T = 5.3$ in this model.

We see then that the assumption of a uniform emitting layer has a bias
towards high densities, which is stronger for the opacity method than
for the radiance method.  A filling factor computed from these values
would be $f=1.5$, while it has been assumed to be $1$ when
computing $\tau_0$ and $E$ from the \citet{avrett08} model: this can
be one of the reasons contributing to our too high observed filling
factor.

This differential bias acts in a surprising way as, due to the $\nE^2$
term in Eq.~(\ref{eq:e}) one would rather expect the bias to be
stronger for $E$ than for $\tau_0$; however, it can be understood by
comparing the effective temperatures for $\tau_0$ and $E$, which are
respectively:
\begin{eqnarray}
  \label{eq:tefft}
  T_{\text{eff},\tau_0} = \frac{\int T(s) \, \nE(s) \, K(T(s))
    \; \ud s}{\int \nE(s) \, K(T(s)) \; \ud s} =  10^{5.38} \unit{K} \\
  \label{eq:teffe}
  T_{\text{eff},E} = \frac{\int T(s) \, \nE^2(s) \, G(T(s)) \;
  \ud s}{\int \nE^2(s) \, G(T(s)) \; \ud s} = 10^{5.40} \unit{K}
\end{eqnarray}
where $K(T) = k_{\nu_0}(T) \, \nSvi / \nE$, while $T(s)$ and $\nE(s)$
are from \citet{avrett08}.  The higher effective temperature for $E$
than for $\tau_0$ means that the bias is more affected by the
respective shapes of the high-temperature wings of $G(T)$ and $K(T)$
than by the exponent of \nE\ in the integrals of Eqns.~(\ref{eq:knu0})
and (\ref{eq:e}).

It can be pointed out here that the difference between the $K(T)$ and
$G(T)$ kernels lies in the fact that $G(T)$ (unlike $K(T)$) not only
takes into account the ionization equilibrium of \svi, but also the
collisions from $i$ to $j$ levels of \svi\ ions.

\subsubsection{Releasing this assumption: a tentative estimate of the
  density gradient around $\log T = 5.3$}
\label{sec:densprof}

\begin{figure}[tp]
  \centering
  \includegraphics[width=\linewidth]{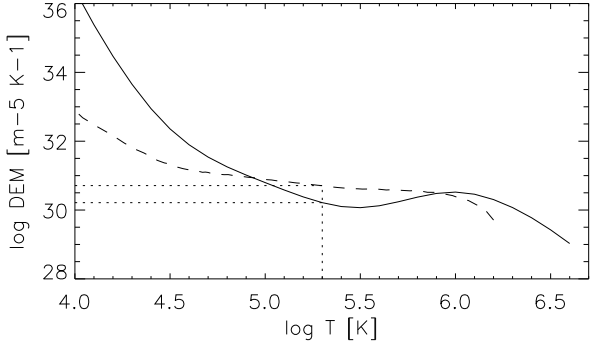}
  \caption{Quiet Sun standard DEM from CHIANTI (plain line) and DEM
    computed from the \citet{avrett08} temperature and density
    profiles.  The dotted lines give the DEMs for $\log T = 5.3$, the
    maximum emission temperature of the $\svi$ lines.}
  \label{fig:dems}
\end{figure}

In Sec. \ref{sec:biasuniform} we have incidentally shown that the
radiance computed with the \citet{avrett08} profiles and the CHIANTI
contribution function $G(T)$ is a factor $3$ higher than the radiance
computed directly by CHIANTI using the standard Quiet Sun DEM (see
Table~\ref{tab:lines}).  This is simply because the DEM computed from
the temperature and density profiles of the \citet{avrett08} model is
different\footnote{The reason for this is that the \citet{avrett08}
  model is determined from theoretical energy balance and needs
  further improvements in order to reproduce the observed DEM (E.\
  Avrett, private communication).} than the CHIANTI DEM, as can be
seen in Fig. \ref{fig:dems}.  In particular, the \citet{avrett08} DEM
is missing the dip around $\log T = 5.5$ that is obtained from most
observations; at $\log T = 5.3$ it is a factor $3$ higher than the
CHIANTI Quiet Sun DEM.

\begin{figure}[tp]
  \centering
  \includegraphics[width=\linewidth]{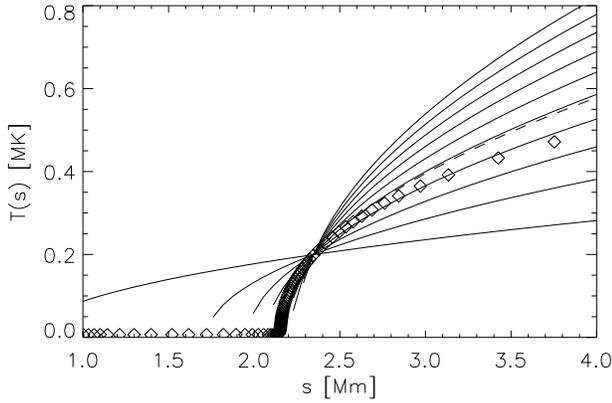}
  \caption{Temperature as a function of altitude in our local
    transition region simple models around $T_0 = 10^{5.3}$ and $s_0 = 2.346
    \unit{Mm}$.  The temperature profile from \citet{avrett08} is
    shown with the diamonds signs, and the simple model with the same
    temperature slope is shown with a dashed line.}
  \label{fig:trmodels}
\end{figure}

We model the upper transition region locally around $\log T_0 = 5.3$
and $s_0 = 2.346 \unit{Mm}$ (chosen because $T(s_0) = T_0$ in the
\citealt{avrett08} model) by a vertically stratified plasma at
pressure $P_0 = 1.91 n_0 k_B T_0$ (we consider a fully ionized coronal
plasma) and:
\begin{equation}
  \frac{T(s)}{T_0} = \frac{n_0}{\nE(s)} =
  \sqrt{\frac{s - s_T}{s_0 - s_T}} \quad \text{for} \quad s > s_T
\end{equation}
These equations were chosen to provide a good approximation of a
transition region, with some symmetry between the opposite curvatures
of the variations of $T$ and $\nE$ with altitude. The parameters of
this model atmosphere are the pressure $P_0$ and $s_T$ (with $s_T <
s_0$), which can be interpreted as the altitude of the base of the
transition region.  Given the constraint $T(s_0) = T_0$ that we
imposed when building the model, with $T_0$ and $s_0$ fixed, $s_T$
actually controls the derivative of $T(s)$ at $s = s_0$:
\begin{equation}
  T'(s_0) = \frac{T_0}{2 (s_0 - s_T)}
  \quad \text{or} \quad
  s_T = s_0 - \frac{T_0}{2T'(s_0)}
\end{equation}
We plot in Fig.~\ref{fig:trmodels} some temperature profiles from this
simple transition region model, for different model parameters
$T'(s_0)$ ($P_0$ only affects the scale of $\nE(s)$).  For the
\citet{avrett08} model, $P_0 = 8.7 \cdot 10^{-3} \unit{Pa}$ and
$T'(s_0) = 0.45 \unit{K \, m^{-1}}$, and the corresponding model
profile is also shown.

\begin{figure}[tp]
  \centering
  \includegraphics[width=\linewidth]{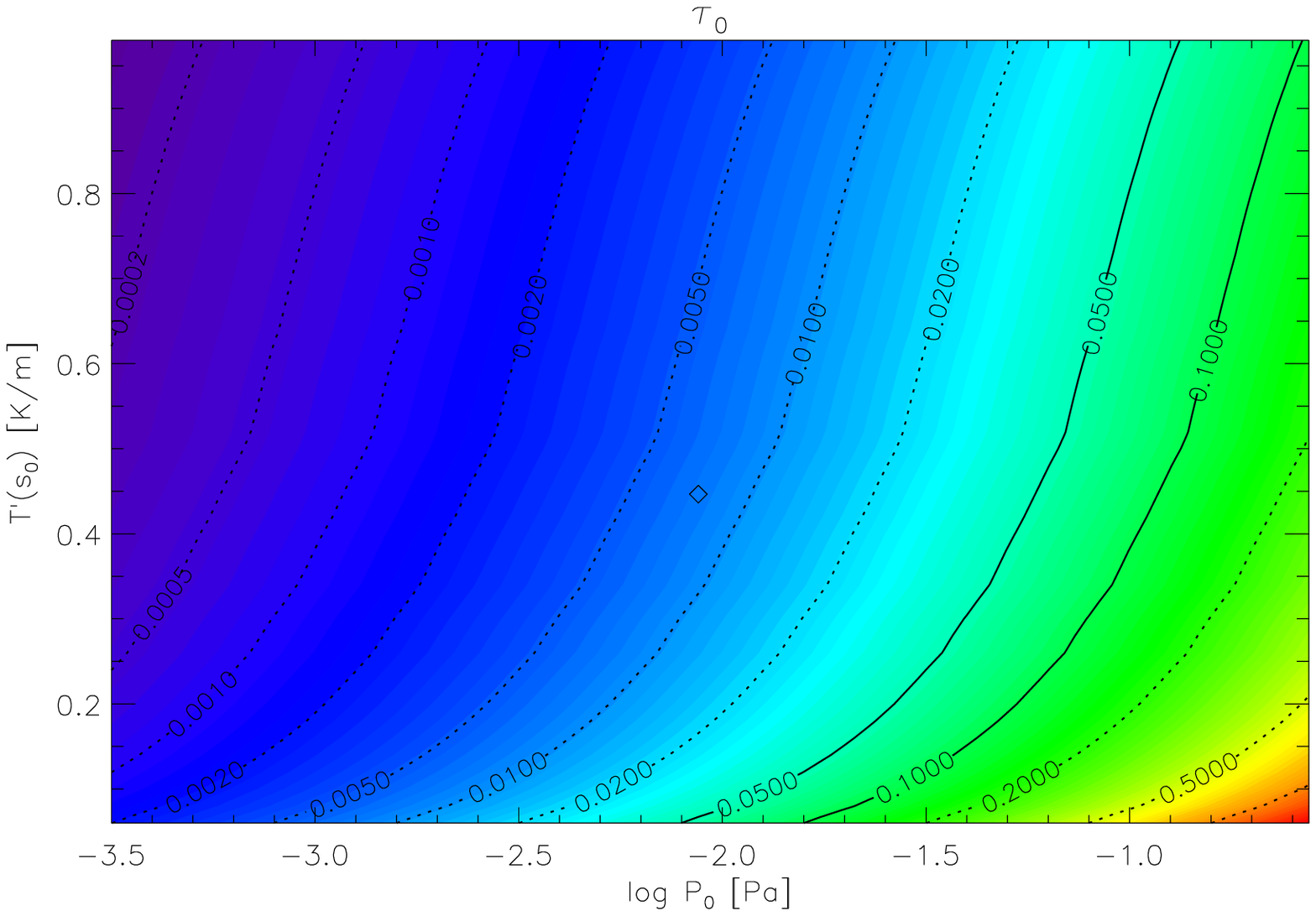}\\
  \includegraphics[width=\linewidth]{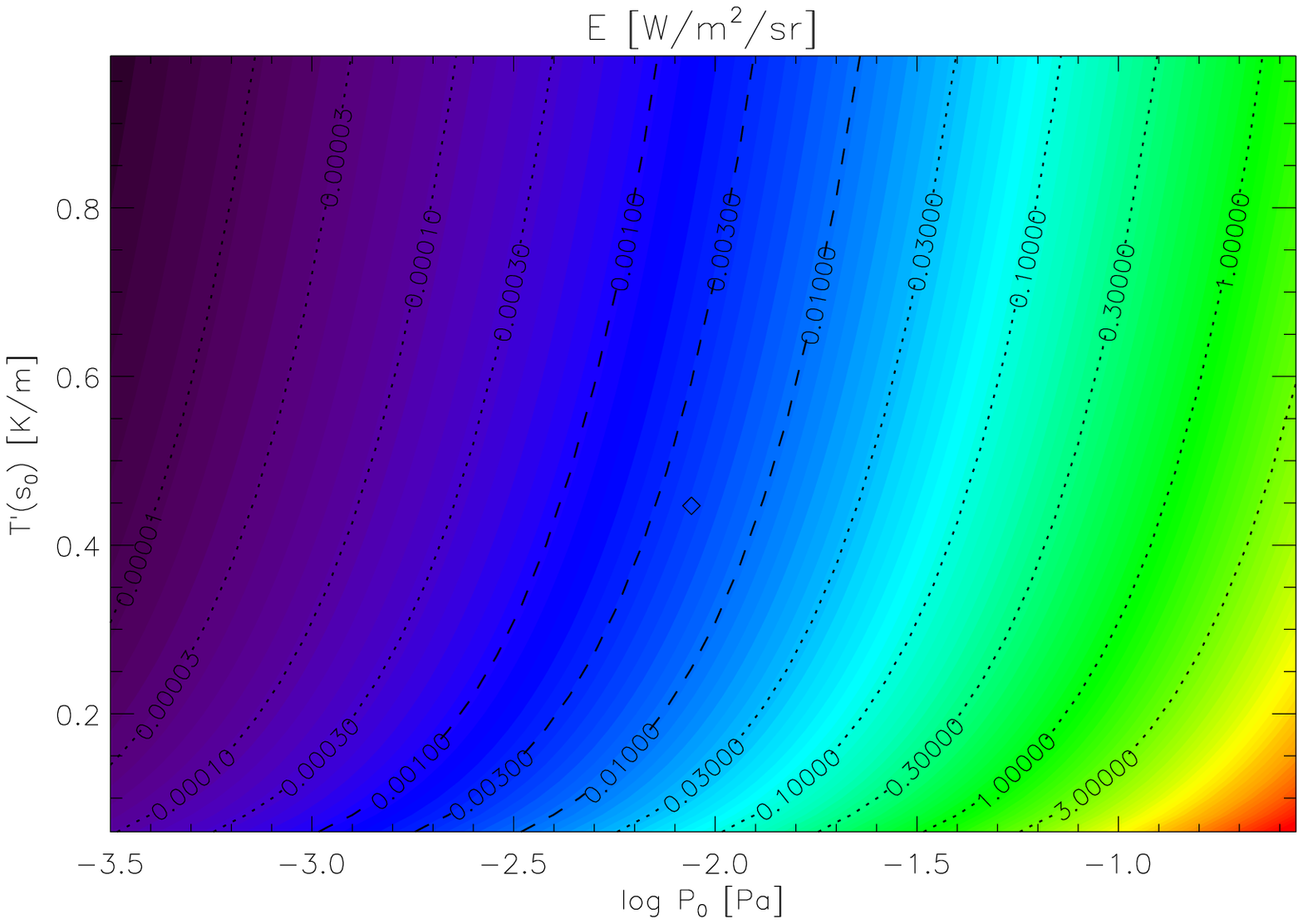}\\
  \includegraphics[width=\linewidth]{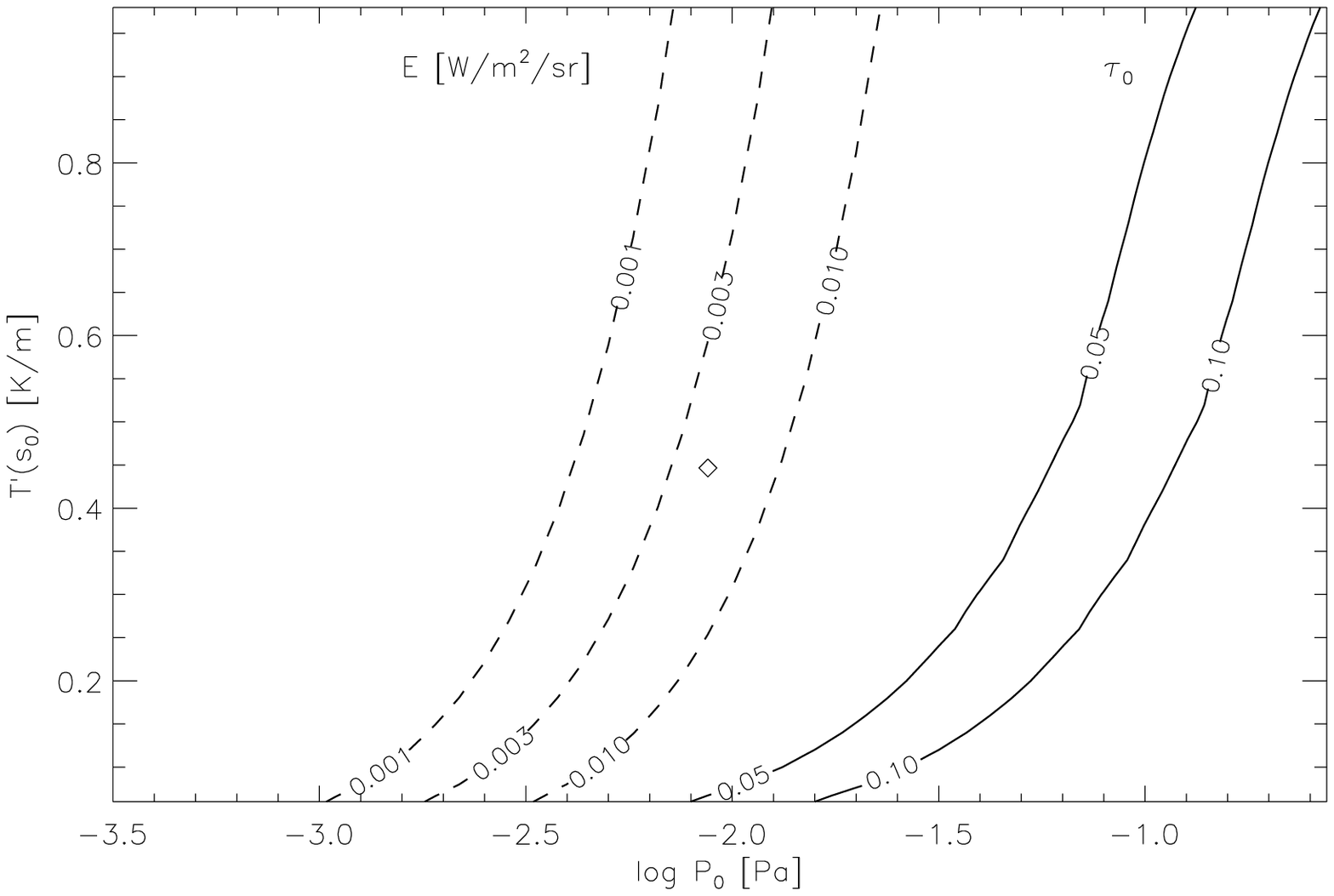}
  \caption{\svii\ opacity $\tau_0$ (top panel) and line radiance $E$
    (middle panel) as a function of model parameters $P_0$ and
    $T'(s_0)$.  The level lines close to our actual observations are
    shown as plain lines for $\tau_0$ and as dashed lines for $E$.
    The bottom panel reproduces these level lines together in the same
    plot.  The parameters $(P_0, T'(s_0))$ estimated from the
    \citet{avrett08} model at $T = T_0$ are shown with the diamond
    sign on each plot.}
  \label{fig:trmodel}
\end{figure}

We propose to use such models along with atomic physics data and the
equations of Sec.~\ref{sec:den} to compute $\tau_0$ and $E$ as a
function of model parameters $P_0$ and $T'(s_0)$, as shown in
Fig.~\ref{fig:trmodel}.  As the slopes of the level lines are
different in the $\tau_0(P_0,T'(s_0))$ and $E(P_0,T'(s_0))$ plots, one
would in theory be able to estimate the parameters $(P_0, T'(s_0))$ of
the best model for the observation of $(\tau_{0,\text{obs}},
E_{\text{obs}})$ by simply finding the crossing between the level
lines $\tau_0(P_0, T'(s_0)) = \tau_{0,\text{obs}}$ and $E(P_0,
T'(s_0)) = E_{\text{obs}}$.

In practice however, the level lines for our observations of $\tau_0$
and $E$ do not intersect in the range of parameters plotted in
Fig.~\ref{fig:trmodel}, corresponding to realistic values of the
parameters.  As a consequence, it is not possible to tell from these
measurements (from a single spectral line, here \svii), what is the
temperature slope and the density of the TR around the formation of
this line.

If we now extend the range of $T'(s_0)$ to unrealistically low values,
a crossing of the level lines can be found below $\log P_0 = -3.5$ and
$T'(s_0) = 5\unit{mK/m}$.  Given the width of $G(T)$ for \svii, this
corresponds to $\Delta s > 20\unit{Mm}$, a value consistent with the
one obtained from Eq.~(\ref{eq:dsff}) and which is also much larger
than expected.

Let us note that \citet{keenan88} derived a much lower \svii\ opacity
value ($\tau_0 = 1.1\,10^{-4}$ at disk center) from a computation
implying the cells of the network model of \citet{gab76}.  However,
while our value of $\tau_0$ seems to be too high, the level lines in
Fig.~\ref{fig:trmodel} show that an opacity value $\tau_0 =
1.1\,10^{-4}$ would be too low: from this figure we expect that a
value compatible with radiance measurements and with realistic values
of the temperature gradient would be in the range $5\,10^{-3}$ to
$10^{-2}$.

\subsection{Anomalous behavior of Na-like ions}
\label{sec:nalike}

Following works such as \citet{dupree72} for Li-like ions,
\citet{judge95} report that standard DEM analysis fails for ions of
the Li and Na isoelectronic sequences; in particular, for \svi\ (which
is Na-like), \citet{delzannag01b} find that the atomic physics models
underestimate the \svii\ line radiance $E$ by a factor $3$.  This
fully explains the difference between our observation of $E$ and the
value computed by CHIANTI (Table~\ref{tab:lines}).  However, this
means also that where $G(T)$ from CHIANTI is used, as in
Eq.~(\ref{eq:e}), it presumably needs to be multiplied by $3$.  As a
result, one can expect $\nERMS$ to be lower by a factor $1.7$,
resulting into a filling factor of 415 (actually worse than our
initial result).

The reasons for the anomalous behavior of these ions for $G(T)$, which
could be linked to the ionization equilibrium or to collisions, are
still unknown.  As a result, it is impossible to tell whether these
reasons also produce an anomalous behavior of these ions for $K(T)$,
hence on our measurements of opacities and on our estimations of
densities: this could again reduce the filling factor.

\subsection{Cell-and-network pattern}
\label{sec:cells}

When analyzing our observations, we have not made the distinction
between the network lanes and the cells of the chromospheric
supergranulation.  Here we try to evaluate the effect of the
supergranular pattern on our measurements, by using a 2D model
emitting layer with a simple ``paddle wheel'' cell-and-network
pattern: in polar coordinates $(r,\theta)$, the emitting layer is
defined by $R_1<r<R_2$; in the emitting layer, the network lanes are
defined by $\theta \in [0, \delta \theta] \mod \Delta \theta$ and the
cells are the other parts of the emitting layer, with $\Delta \theta$
the pattern angular cell size (an integer fraction of $2\pi$) and
$\delta \theta$ the network lane angular size.  The network lanes and
cells are characterized by different (but uniform) source functions
$S$, densities $\nSvi$ and absorption coefficients $k_{\nu_0}$.  We
then solve the radiative transfer equations for $\lambda_0$ along rays
coming from infinity through the emitting layer to the
observer.

\begin{figure}[tp]
  \centering
  \includegraphics[width=\linewidth]{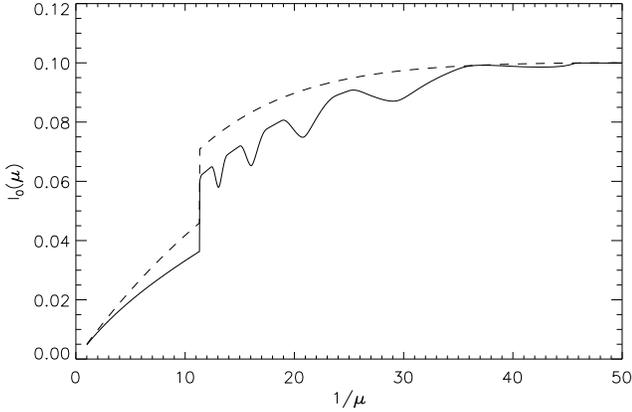}
  \caption{Average spectral radiance at line center $I_0$ as a
    function of $1/\mu$ for a uniform layer (dashed line) and for a
    model layer with a simple cell-and-network pattern (plain line).
    Both models have the same average opacity and source function.
    The factor-$2$ jump at $1/\mu = 11.3$ corresponds to the limb of
    the opaque solar disk; the reference radius used to compute $\mu$
    corresponds to the middle of the emitting layer.  The oscillations
    are artefacts of the averaging process.}
  \label{fig:rouhness2}
\end{figure}

As the opacity is obtained by a simple integration of $k_{\nu_0}
\nSvii$, the average line-of-sight opacity $t_0$ as a function of
$\mu$ for the ``paddle-wheel'' pattern is the same as for a uniform
layer with the same average $k_{\nu_0} \nSvii$. However, as seen in
Fig.~\ref{fig:rouhness2}, still for the same average $S$ and
$k_{\nu_0} \nSvii$, the effect of opacity (a decrease in intensity) is
higher in the ``paddle-wheel'' case, in particular for intermediate
values of $1/\mu$.  As a result, neglecting the cell-and-network
pattern of the real TR leads to overestimating the opacity when
using method A.

\subsection{Roughness and fine structure}
\label{sec:roughness}

In order to explain the high values of opacity (as derived from their
method A), \citet{dumont83} introduce the concept of roughness of the
TR: as the TR plasma is not perfectly vertically stratified (there is
some horizontal variation), method A leads to an overestimated value
of $\tau_0$.  This could reconcile the values obtained following our
application of methods A and B.

\begin{figure}[tp]
  \centering
  \includegraphics[width=\linewidth]{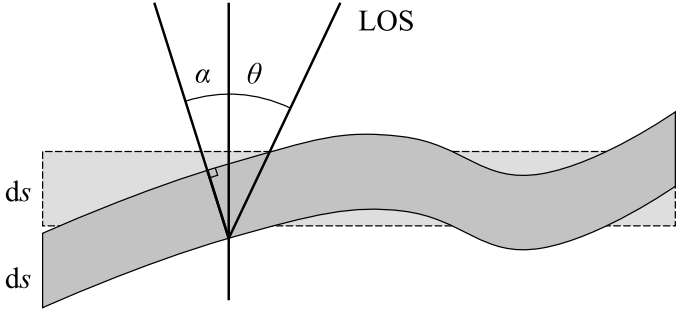}
  \caption{Geometry of a TR layer (plain contour), displaced from its
    average position (dashed contour) while retaining its original
    vertical thickness $\ud s$, and locally forming an angle $\alpha$
    with the average layer.  The line-of-sight (LOS) forms an angle
    $\theta$ to the vertical (normal to the average layer).}
  \label{fig:roughtr}
\end{figure}

We model the roughness of the transition region by incompressible
vertical displacements of any given layer (at given optical depth)
from its average vertical position, in the geometry shown in
Fig.~\ref{fig:roughtr}.  The layer then forms an angle $\alpha$ with
the horizontal and has still the same vertical thickness $\ud s$;
the thickness along the LOS is $\ud s \cos \alpha / \cos (\theta +
\alpha)$, as can be deduced from Fig.~\ref{fig:roughtr}.

If we assume that $\theta + \alpha$ remains sufficiently small for the
plane-parallel approximation to hold (and so that the LOS crosses one
given layer only once), the opacity is
\begin{eqnarray}
  t_0 = \int \nE(s) K(T(s)) \frac{\cos \alpha \ud s}{\cos (\theta +
    \alpha)} 
\end{eqnarray}

The angle $\alpha$ is a random variable, with some given distribution
${\Pr} (\alpha)$.  We compute the average of $t_0$ as a
function of $\theta$ and of $\Pr(\alpha)$:
\begin{eqnarray}
  \left\langle t_0 (\theta, \Pr(\alpha)) \right\rangle_\alpha & =
  {\displaystyle \iint \nE(s)
    K(T(s)) \frac{\cos \alpha \ud s}{\cos (\theta + \alpha)}
    \Pr(\alpha) \ud \alpha} \\  
  & = {\displaystyle \frac{\tau_0}{\mu} \left\langle \frac{\cos \theta
        \cos \alpha}{\cos (\theta + \alpha)} \right\rangle}_\alpha \equiv
  {\displaystyle \frac{\tau_0}{\mu} \beta(\theta,\Pr(\alpha))}
\end{eqnarray}
The opacity $t_0 = \tau_0 / \mu$ is corrected by the factor $\beta
(\theta, \Pr(\alpha))$ defined in the previous equation.  We recover
$\beta = 1$ for $\Pr(\alpha) = \delta(\alpha)$, i.e., when there is no
roughness.

We immediately see that $\beta = 1$ for $\theta = 0$, for any
$\Pr(\alpha)$: roughness (as modelled here by incompressible vertical
displacements) does not change the optical thickness at disk center.
Nevertheless, the \textit{estimate} of optical thickness at disk
center from observations in Sec.~\ref{sec:methoda} \citep[method A
of][]{dumont83} is affected by this roughness effect.

\begin{figure}[tp]
  \centering
  \includegraphics[width=\linewidth]{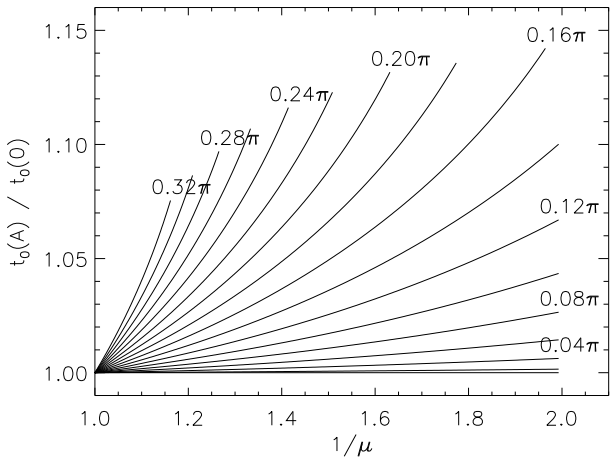}
  \caption{Multiplicative coefficient to $t_0$ due to roughness, for
    different roughness parameters $A$.}
  \label{fig:roughness}
\end{figure}

Coming back to $\langle t_0 \rangle$, we take $\Pr(\alpha) = \cos^2
(\pi \alpha / 2A) / A$, and we compute $\beta$ numerically ($A$
represents the width of $\Pr(\alpha)$ and can be thought as a
quantitative measurement of the roughness).  The results, shown in
Fig.~\ref{fig:roughness}, indicate for example that the modelled
roughness with $A=\pi/5$ increases the opacity by $9\%$ at
$1/\mu=1.5$ (corresponding to $\theta=45$\textdegree).  This is a
significant effect, and we can evaluate its influence on the estimate
of $\tau_0$ in Sec.~\ref{sec:methoda}: in the theoretical profiles of
$I_0(\mu)$ and $E(\mu)$ (Eq.~\ref{eq:iprof}--\ref{eq:eprof}), $\tau_0
/ \mu$ needs to be replaced by $\tau_0 / \mu \times \beta$.  As $\beta
> 1$ for a rough corona, this means that the value of $\tau_0$
determined from the fit of observed radiances to
Eq.~(\ref{eq:iprof})--(\ref{eq:eprof}) is overestimated by a factor
corresponding approximately to the mean value of $\beta$ on the
fitting range.

In this way, we have given a quantitative value for the overestimation
factor of $\tau_0$ by the method of Sec.~\ref{sec:methoda}, thus
extending the qualitative discussion on roughness found in
\citet{dumont83}.  This factor, of the order of $1.1$ may seem modest,
but one needs to remember that the fit for obtaining $\tau_0$ in
Sec.~\ref{sec:methoda} was done on a wide range ($1/\mu$ from $1$ to
$5$, or $\theta$ from $0$ to $78$ degrees) that our roughness model
cannot reproduce entirely\footnote{For high values of the
  $\Pr(\alpha)$ width $A$, the correction factor $\beta$ cannot be
  computed for high values of $1/\mu$ (high angles $\theta$) because
  the values of $\alpha$ in the wings of $\Pr(\alpha)$ fall in the
  range where $|\theta + \alpha| \nll \pi/2$: the plane-parallel
  approximation is not valid anymore.  This explains the limited range
  of the $\beta(1/\mu)$ curves in Fig.~\ref{fig:roughness}.}.

One can think of different roughness models representing the strong
inhomogeneity of the TR, for instance with a different and very
peculiar roughness model \citet{pecker88} obtain an overestimation
factor of more than 10 under some conditions.  This means that our
values of $\tau_0$ may need to be decreased by a large factor due to a
roughness effect.

Roughness models can be seen as simplified models of the fine
structure of the TR, which is known to be heterogeneous at small
scales. Indeed, in addition to the chromospheric network pattern that
we have already modelled in Sec.~\ref{sec:cells}, the TR contains
parts of different structures, with different plasma properties, like
the base of large loops and coronal funnels, smaller loops
\citep{dow86,peter01}, and spicules.  Furthermore, the loops
themselves are likely to be composed of strands, which can be heated
independently \citep{car04,parenti06}.  The magnetic field in these
structures inhibits perpendicular transport, and as a consequence the
horizontal inhomogeneities are not smoothed out efficiently.

\section{Conclusion}

We have first derived the average electron density in the TR from the
opacity $\tau_0$ of the \svii\ line, obtained by three different
methods from observations of the full Sun: center-to-limb variation of
radiance, center-to-limb ratios of radiance and line width, and
radiance ratio of the $93.3$--$94.4 \unit{nm}$ doublet.  Assuming a
spherically symmetric plane-parallel layer of constant source
function, we find a \svii\ opacity of the order of $0.05$.  The
derived average electron density is of the order of $2.4 \cdot 10^{16}
\unit{m^{-3}}$.

We have then used the line radiance (by an EM method) in
order to get the RMS average electron density in the \svii-emitting
region: we obtain $2.0 \cdot 10^{15} \unit{m^{-3}}$.  This corresponds
to a total pressure of $10^{-2} \unit{Pa}$, slightly higher than the
range of pressures found by \citet{dumont83} ($1.3$ to $6.3 \cdot
10^{-3} \unit{Pa}$, as deduced from their Sec. 4.2), but lower than
the value given in \cite{mariska93book} ($2 \cdot 10^{-2} \unit{Pa}$).

The average electron densities obtained from these methods (opacity on
one hand, radiance on the other hand) are incompatible, as can be seen
either from a direct comparison of the values of $\nEav$ and $\nEcav$
for a given thickness $\Delta s$ of a uniform emitting layer, or by
computing the $\Delta s$ that would reconcile the measurements of
$\nEav \, \Delta s$ and $\nEcav \, \Delta s$. Furthermore, we have
seen that the density obtained from the opacity method is also
incompatible with standard DEMs of the Quiet Sun (see
Sec.~\ref{sec:ne2}) and with semi-empirical models of the temperature
and density profiles in the TR (see Sec.~\ref{sec:densprof}).

We investigated several possible sources of biases in the
determination of $\tau_0$: the approximation of a constant temperature
in the \svi\ emitting layer, the anomalous behavior of the \svi\ ion,
the chromospheric network pattern, and the roughness of the TR.  Some
of these could help explain partly the discrepancy between the average
densities deduced from opacities and from radiances, but there is
still a long way to go to fully understand this discrepancy and to
reconcile the measurements.  At this stage, we can only encourage
colleagues to look for similar discrepancies in lines formed around
$\log T = 5.3$ (like \ion{C}{iv} and \ion{O}{vi}), Na-like and not
Na-like, and to repeat similar \svi\ center-to-limb measurements.

In Sec.~\ref{sec:densprof} we have tried to combine opacity and
radiance information to compute the gradient of temperature. This
appeared to be impossible (if restricting ourselves to a realistic
range of parameters) because of the above-mentioned incompatibility.
We have estimated that a value $\tau_0$ of the \svii\ opacity
compatible with radiance measurements and with realistic values of the
temperature gradient would be in the range $5\,10^{-3}$ to $10^{-2}$.

In spite of the difficulties we met, we still think that the
combination of opacity and radiance information should be a powerful
tool for investigating the thermodynamic properties and the fine
structure of the TR.  For instance the excess opacity derived from
observations and a plane-parallel model could be used to evaluate
models of roughness and fine structure of the TR.  Clearly, progress
in modelling the radiative output of the complex structure of the TR
needs to be done in order to achieve this.

\begin{acknowledgements}
  The authors thank G.\ del Zanna, E.\ H.\ Avrett and Ph.\ Lemaire for
  interesting discussions and the anonymous referee for suggestions
  concerning this paper.  The SUMER project is supported by DLR, CNES,
  NASA and the ESA PRODEX Programme (Swiss contribution). SoHO is a
  project of international cooperation between ESA and NASA. Data was
  provided by the MEDOC data center at IAS, Orsay.  EB thanks CNES for
  financial support, and the ISSI group on Coronal Heating (S.\
  Parenti). CHIANTI is a collaborative project involving the NRL
  (USA), RAL (UK), MSSL (UK), the Universities of Florence (Italy) and
  Cambridge (UK), and George Mason University (USA).
\end{acknowledgements}

\appendix

\section{About the filling factor}
\label{sec:ff}

In this paper we have defined the filling factor as 
\begin{equation}
  f = \frac{\nEav^2}\nEcav
\end{equation}
while it is usually obtained, from solar observations
\citep[e.g.][]{judge00,klicar01}, from
\begin{equation}
  \label{eq:ff1}
  f=\frac{EM}{\Delta s \, n_0^2}
\end{equation}
where $EM$ is the emission measure, $\Delta s$ is the thickness of the
plasma layer and $n_0$ is the electron density (usually determined
from line ratios) \emph{in the non-void parts of the plasma layer}.

It may seem surprising that the $EM$ is at the numerator of this
second expression, while it provides an estimate for $\nEcav$ which
appears at the denominator of the first expression.  However, we can
show that these both expressions, despite looking very different, give
actually the same result for a given plasma.

We take a plasma with a differential distribution $\xi(\nE,T)$ for the
density and temperature: $\xi(\nE,T) \; \ud \nE \; \ud T$ is the
proportion of any given volume occupied by plasma at a density between
$\nE$ and $\nE + \ud \nE$ and a temperature between $T$ and $T + \ud
T$.

The contributions to the line radiance $E$ and to the opacity at line
center $\tau_0$ from a volume $V$ with this plasma distribution are
\begin{align}
  \frac E V = \iint \nE^2 G(\nE,T) \, \xi(\nE,T) \; \ud \nE \; \ud T \\
  \frac {\tau_0} V = \iint \nE K(\nE,T) \, \xi(\nE,T) \; \ud \nE \;
  \ud T
\end{align}
with the notations of our article.

The usual assumption \citep[e.g.][]{judge00} is that $G(T,\nE)$
``selects'' a narrow range of temperatures around $T=T\imax$ and does
not depend on $\nE$, i.e., $G(\nE,T) \approx \tilde G (T\imax) \,
\delta (T - T\imax)$.  Similarly, we can consider that $K(\nE,T)
\approx \tilde K (T\imax) \, \delta (T - T\imax)$.  Then
\begin{align}
  \frac E V &\approx \tilde G(T\imax) \int \nE^2 \, \xi(\nE,T\imax) \;
  \ud \nE = \tilde G(T\imax) \,
  \nEcav_{T=T\imax} \\
  \frac {\tau_0} V &\approx \tilde K(T\imax) \int \nE \,
  \xi(\nE,T\imax) \; \ud \nE = \tilde K(T\imax) \, \nEav_{T=T\imax}
\end{align}

The line ratio $R_{ij} = E_i / E_j$ is, following \citet{judge00} and
with the assumption $G(\nE,T) = \hat G(\nE,T\imax) \, \delta (T - T\imax)$:
\begin{align}
  R_{ij} = \frac{E_i}{E_j} &= \frac{ \iint \nE^2 G_i (\nE,T) \,
    \xi(\nE,T) \; \ud \nE \; \ud T }{ \iint \nE^2 G_j (\nE,T) \,
    \xi(\nE,T) \; \ud \nE \; \ud T
  } \\
  & \approx \frac{ \int \nE^2 \hat G_i (\nE,T\imax) \, \xi(\nE,T\imax)
    \; \ud \nE }{ \int \nE^2 \hat G_j (\nE,T\imax) \, \xi(\nE,T\imax)
    \; \ud \nE
  } 
\end{align}
When homogeneity is assumed, i.e., $\xi(\nE,T) = \delta(\nE - n_0) \,
\tilde\xi(T)$, this becomes
\begin{equation}
  R_{ij} \approx \frac{n_0^2 G_i (n_0, T\imax) \,
    \tilde\xi(T\imax)}{n_0^2 G_j (n_0,
    T\imax) \, \tilde\xi(T\imax)} = \frac{G_i (n_0,
    T\imax)}{G_j (n_0, 
    T\imax)} \equiv g_{ij}(n_0)
\end{equation}
and inverting this function allows to recover $n_0$ from the observed
value of $R_{ij}$.

The fundamental point is that $R_{ij}$ does not depend on the
proportion $f$ (the filling factor) of the volume actually occupied by
the plasma: $n_0$ is the density in the non-void region only.  For
example, for $\xi_f(\nE,T) $ defined by $f \delta(\nE-n_0) +
(1-f)\delta(\nE)$, the line ratio $R_{ij}$ is $g_{ij}(n_0)$ which is
independent on $f$, while $\nEcav_{T=T\imax}$ determined from $E/V$
would be $f n_0^2$ and $\nEav_{T=T\imax}$ determined from $\tau_0/V$
would be $f n_0$.  One can see in this case that $f$ can
(equivalently) either be recovered from
\begin{equation}
  \frac{\nEcav_{T=T\imax}}{(n_0)^2} = \frac{(f n_0^2)}{(n_0)^2} = f
\end{equation}
(corresponding to \citealt{judge00}) or from
\begin{equation}
  \frac{\nEav_{T=T\imax}^2}{\nEcav_{T=T\imax}} = \frac{(f n_0)^2}{(f
    n_0^2)} = f
\end{equation}
(corresponding to our method).


\begin{thebibliography}{23}
\expandafter\ifx\csname natexlab\endcsname\relax\def\natexlab#1{#1}\fi

\bibitem[{{Avrett} \& {Loeser}(2008)}]{avrett08}
{Avrett}, E.~H. \& {Loeser}, R. 2008, Astrophys. J. Suppl. Ser., 175, 229

\bibitem[{{Buchlin} {et~al.}(2006){Buchlin}, {Vial}, \& {Lemaire}}]{buc06}
{Buchlin}, E., {Vial}, J.-C., \& {Lemaire}, P. 2006, Astron. Astrophys., 451,
  1091

\bibitem[{{Cargill} \& {Klimchuk}(2004)}]{car04}
{Cargill}, P.~J. \& {Klimchuk}, J.~A. 2004, Astrophys. J., 605, 911

\bibitem[{{Curdt} {et~al.}(2001){Curdt}, {Brekke}, {Feldman}, {Wilhelm},
  {Dwivedi}, {Sch{\" u}hle}, \& {Lemaire}}]{cur01}
{Curdt}, W., {Brekke}, P., {Feldman}, U., {et~al.} 2001, Astron. Astrophys.,
  375, 591

\bibitem[{{Del Zanna} {et~al.}(2001){Del Zanna}, {Bromage}, \&
  {Mason}}]{delzannag01b}
{Del Zanna}, G., {Bromage}, B. J.~I., \& {Mason}, H.~E. 2001, in American
  Institute of Physics Conference Series, Vol. 598, Joint SOHO/ACE workshop
  ''Solar and Galactic Composition'', ed. R.~F. {Wimmer-Schweingruber}, 59--64

\bibitem[{{Dere} {et~al.}(1997){Dere}, {Landi}, {Mason}, {Monsignori Fossi}, \&
  {Young}}]{chianti}
{Dere}, K.~P., {Landi}, E., {Mason}, H.~E., {Monsignori Fossi}, B.~C., \&
  {Young}, P. 1997, Astron. Astrophys. Suppl. Ser., 125, 149

\bibitem[{{Dowdy} {et~al.}(1986){Dowdy}, {Rabin}, \& {Moore}}]{dow86}
{Dowdy}, J.~F., {Rabin}, D., \& {Moore}, R.~L. 1986, Sol. Phys., 105, 35

\bibitem[{{Dumont} {et~al.}(1983){Dumont}, {Pecker}, {Mouradian}, {Vial}, \&
  {Chipman}}]{dumont83}
{Dumont}, S., {Pecker}, J.-C., {Mouradian}, Z., {Vial}, J.-C., \& {Chipman}, E.
  1983, Sol. Phys., 83, 27

\bibitem[{{Dupree}(1972)}]{dupree72}
{Dupree}, A.~K. 1972, Astrophys. J., 178, 527

\bibitem[{{Gabriel}(1976)}]{gab76}
{Gabriel}, A. 1976, Royal Society of London Philosophical Transactions Series
  A, 281, 339

\bibitem[{{Judge}(2000)}]{judge00}
{Judge}, P.~G. 2000, Astrophys. J., 531, 585

\bibitem[{{Judge} {et~al.}(1995){Judge}, {Woods}, {Brekke}, \&
  {Rottman}}]{judge95}
{Judge}, P.~G., {Woods}, T.~N., {Brekke}, P., \& {Rottman}, G.~J. 1995,
  Astrophys. J., 455, L85+

\bibitem[{{Keenan}(1988)}]{keenan88}
{Keenan}, F.~P. 1988, Sol. Phys., 116, 279

\bibitem[{{Klimchuk} \& {Cargill}(2001)}]{klicar01}
{Klimchuk}, J.~A. \& {Cargill}, P.~J. 2001, Astrophys. J., 553, 440

\bibitem[{{Landi} {et~al.}(2006){Landi}, {Del Zanna}, {Young}, {Dere}, {Mason},
  \& {Landini}}]{landie06}
{Landi}, E., {Del Zanna}, G., {Young}, P.~R., {et~al.} 2006, Astrophys. J.
  Suppl. Ser., 162, 261

\bibitem[{Mariska(1993)}]{mariska93book}
Mariska, J.~T. 1993, The Solar Transition Region (Cambridge University Press)

\bibitem[{{Mason}(1998)}]{mason98lnp}
{Mason}, H.~E. 1998, in Lecture Notes in Physics, Berlin Springer Verlag, Vol.
  507, Space Solar Physics: Theoretical and Observational Issues in the Context
  of the SOHO Mission, ed. J.~C. {Vial}, K.~{Bocchialini}, \& P.~{Boumier},
  143--+

\bibitem[{{Parenti} {et~al.}(2006){Parenti}, {Buchlin}, {Cargill}, {Galtier},
  \& {Vial}}]{parenti06}
{Parenti}, S., {Buchlin}, E., {Cargill}, P.~J., {Galtier}, S., \& {Vial}, J.-C.
  2006, Astrophys. J., 651, 1219

\bibitem[{{Pecker} {et~al.}(1988){Pecker}, {Dumont}, \& {Mouradian}}]{pecker88}
{Pecker}, J.-C., {Dumont}, S., \& {Mouradian}, Z. 1988, Astron. Astrophys.,
  196, 269

\bibitem[{{Peter}(1999)}]{peter99}
{Peter}, H. 1999, Astrophys. J., 516, 490

\bibitem[{{Peter}(2001)}]{peter01}
{Peter}, H. 2001, Astron. Astrophys., 374, 1108

\bibitem[{{Peter} \& {Judge}(1999)}]{peter99b}
{Peter}, H. \& {Judge}, P.~G. 1999, Astrophys. J., 522, 1148

\bibitem[{{Wilhelm} {et~al.}(1995){Wilhelm}, {Curdt}, {Marsch}, {Sch{\" u}hle},
  {Lemaire}, {Gabriel}, {Vial}, {Grewing}, {Huber}, {Jordan}, {Poland},
  {Thomas}, {Kuhne}, {Timothy}, {Hassler}, \& {Siegmund}}]{sumer}
{Wilhelm}, K., {Curdt}, W., {Marsch}, E., {et~al.} 1995, Sol. Phys., 162, 189

\end{thebibliography}
\end{document}